\documentclass[letterpaper,11pt]{article}
\pdfoutput=1 

\usepackage{jheppub} 
            
\usepackage{bm,amssymb,slashed,graphicx,multirow,soul,mathtools,xspace,array}  
\usepackage{float}                   
\allowdisplaybreaks
\usepackage{ bbold }
\usepackage{subfigure}

\newcommand{\todo}[1]{{\color{red} \ifmmode\else[todo]\fi #1}}
\usepackage[usenames,dvipsnames]{xcolor}
     \definecolor{hgreen}{rgb}{0,.3,0}
      \definecolor{darkgreen}{rgb}{0.3,.8,0.2}
     \definecolor{hred}{rgb}{.3,0,0}
     \definecolor{hblue}{rgb}{0,0,.3}
     \definecolor{LightGray}{gray}{0.95}
     
\usepackage{hyperref}

\usepackage{mathrsfs}

 \usepackage{etoolbox}
    \makeatletter
    \patchcmd{\maketitle}{\@fpheader}{}{}{}
    \makeatother

\newcommand\snowmass{
\begin{center}
  \rule[-0.2in]{\hsize}{0.01in}\\
  \rule{\hsize}{0.01in}\\
  \vskip 0.1in
  Submitted to the Proceedings of the US Community Study\\ 
  on the Future of Particle Physics (Snowmass 2021)\\
  \rule{\hsize}{0.01in}\\
  \rule[+0.2in]{\hsize}{0.01in}\\[-2em]
\end{center}
}

\def\beq{\begin{equation}}
\def\eeq{\end{equation}}

\def\vect#1{\boldsymbol{#1}}

\title{Snowmass White Paper: Effective Field Theories for Dark Matter Phenomenology}

\author[a]{Matthew Baumgart,}
\author[b]{Fady Bishara,}
\author[c]{Joachim~Brod,}
\author[d]{Timothy Cohen,}
\author[e]{A. Liam Fitzpatrick,}
\author[f]{Martin Gorbahn,}
\author[g]{Ulserik Moldanazarova,}
\author[h]{Matthew Reece,}
\author[i]{Nicholas L. Rodd,}
\author[j]{Mikhail P. Solon,}
\author[k]{Robert Szafron,}
\author[l]{Zhengkang Zhang,}
\author[c]{Jure Zupan}

\affiliation[a]{Department of Physics, Arizona State University, Tempe, AZ 85287, USA}
\affiliation[b]{Deutsches Elektronen-Synchrotron DESY, Notkestr. 85, 22607 Hamburg, Germany}
\affiliation[c]{Department of Physics, University of Cincinnati, Cincinnati, Ohio 45221, USA}
\affiliation[d]{Institute for Fundamental Science, University of Oregon, Eugene, OR 97403, USA}
\affiliation[e]{Department of Physics, Boston University, Boston, MA 02215, USA}
\affiliation[f]{Department of Mathematical Sciences, University of Liverpool, Liverpool L69 3BX, United Kingdom}
\affiliation[g]{Faculty of Physics and Technology, Karaganda Buketov University, 100028 Karaganda, Kazakhstan}
\affiliation[h]{Department of Physics, Harvard University, Cambridge, MA, 02138, USA}
\affiliation[i]{Theoretical Physics Department, CERN, 1 Esplanade des Particules, CH-1211 Geneva 23, Switzerland}
\affiliation[j]{Mani L. Bhaumik Institute for Theoretical Physics, University of California at Los Angeles, Los Angeles, CA 90095, USA}
\affiliation[k]{Department of Physics, Brookhaven National Laboratory, Upton, N.Y., 11973, USA}
\affiliation[l]{Department of Physics, University of California, Santa Barbara, CA 93106, USA}

\date{\today}

\abstract{The quest to discover the nature of dark matter continues to drive many of the experimental and observational frontiers in particle physics, astronomy, and cosmology. While there are no definitive signatures to date, there exists a rich ecosystem of experiments searching for signals for a broad class of dark matter models, at different epochs of cosmic history, and through a variety of processes with different characteristic energy scales. Given the multitude of candidates and search strategies, effective field theory has been an important tool for parametrizing the possible interactions between dark matter and Standard Model probes, for quantifying and improving model-independent uncertainties, and for robust estimation of detection rates in the presence of large perturbative corrections. This white paper summarizes a wide range of effective field theory applications for connecting dark matter theories to experiments.
}

\begin{document}

\snowmass

 \maketitle
 
 \vfill

\pagebreak
\section{Executive Summary}

Effective Field Theory (EFT) techniques enable precise calculations of dark matter (DM) signals, and have already had a significant impact on the experimental searches in a number of cases, such as predicting DM annihilation rates relevant for imaging atmospheric Cherenkov telescopes, as well as DM scattering rates relevant for flagship direct detection experiments.  The fixed order calculation of annihilation cross sections for heavy, TeV-scale, weakly-interacting massive particle (WIMP) dark matter
into photons 
suffers from 
large Sudakov double logarithms.  These were
resummed using Soft Collinear Effective Theory to next-to-leading logarithmic (NLL) accuracy~\cite{Baumgart:2018yed}, reducing theoretical uncertainties from a factor of a few  
to the level of $\sim$5\%.
The improved predictions are being used by members of the HESS collaboration for interpreting their results~\cite{Rinchiuso:2018ajn}, and are similarly incorporated into the ongoing analysis by the VERITAS telescope. 
The inclusion of the EFT effects is crucial: the leading order cross section for the Higgsino is just within the discovery reach of CTA~\cite{Rinchiuso:2020skh} (assuming a canonical Milky Way dark matter distribution), and whether this well-motivated WIMP candidate can be discovered or excluded in the coming years hinges on where higher order corrections move the leading order prediction.
 In direct detection the use of  EFTs  allows for a model independent comparison of experiments. Because the momentum exchange in DM scattering on nuclei is small, the DM EFTs capture large classes of DM models, as long as mediators are heavier than a GeV. The direct detection EFT framework, either based on DM interactions with nucleons, or on DM couplings to quarks and gluons, was already adopted by a large number of experimental collaborations when interpreting their results~\cite{XENON:2017fdd,PandaX-II:2018woa,CRESST:2018vwt,DarkSide-50:2020swd,CDEX:2020tkb,LUX:2021ksq,LUX:2020oan,IceCube:2021tdt}.

This white paper also highlights a variety of other uses of EFT that further our understanding of DM phenomenology.   High precision predictions for direct detection of heavy WIMPs were obtained through the use of a heavy particle EFT, while potential non-relativistic EFT was used to perform precise thermal relic abundance calculations for heavy WIMPs in the regime where Sommerfeld enhancements are important. Similarly, the implications of DM self-interactions for structure formation can be understood using nonrelativistic EFT. Additionally, new direct detection techniques for light DM with mass below a GeV rely on condensed matter effects that can be well described using EFTs. Finally, the collider searches for DM production have been interpreted using simplified models for the better part of the past decade.  

Given these successes, it is clear that continued development of EFT tools for dark matter will be important for interpreting signals from current and future experiments, especially in the event of a discovery.

\section{Heavy DM: Direct Detection}
\label{sec:HeavyDM}

In a large class of models, the Standard Model is extended at low energies by one or a few particles transforming under definite representations of $SU(2)_W \times U(1)_Y$ with masses much heavier than the electroweak scale, $M \gg m_W$. This is the case for a thermal relic electroweak triplet (wino) or doublet (Higgsino), and is consistent with null results at the LHC and at direct detection experiments. For such models, we can apply a heavy particle formalism familiar from Heavy Quark Effective Theory~\cite{Isgur:1989vq,Caswell:1985ui,Eichten:1989zv} to develop a systematic framework for describing the interactions of such heavy WIMPs with Standard Model particles~\cite{Hill:2011be,Hill:2013hoa}.

Analogous to the case of heavy quarks, heavy WIMP symmetry emerges in the $M \gg m_W$ limit. Spin-independent WIMP-nucleon scattering cross sections become universal for given WIMP electroweak quantum numbers, independent of the details of the UV completion. For example, the cross section does not depend on whether the DM is a scalar or fermion, or whether it is fundamental or composite in nature. This universality can be parametrized systematically in the $1/M$ expansion:
\begin{align}\label{eq:HWlagrangian}
  {\cal L} &= \overline{h}_v \bigg\{ iv\cdot D - \delta m - {D_\perp^2\over 2M}
  + c_H {H^\dagger H\over M}
  + c_{W1}  {\sigma^{\mu\nu} W_{\mu\nu} \over M}
  + c_{W2}  {\epsilon^{\mu\nu\rho\sigma} \sigma_{\mu\nu} W_{\rho\sigma} \over M}
  +  \dots \bigg\} h_v \,.  
\end{align}
This Lagrangian specifies the interactions of $h_v$, a heavy multiplet of self-conjugate particles with arbitrary spin and transforming under irreducible representations of electroweak $SU(2)_W \times U(1)_Y$.\footnote{The timelike unit vector $v^\mu$ defines the heavy WIMP velocity, $D^\mu$ and $W^{\mu \nu}$ are the usual covariant derivative and field strengths, and $D_\perp^\mu  = D^\mu - v^\mu v \cdot D$. See, {\it e.g.}, Refs.~\cite{Hill:2014yka,Chen:2018uqz} for more details.} The coupling $c_H$ gives the leading correction to the universal spin-independent cross-section in the heavy WIMP limit, and encodes ultraviolet physics, which can be determined by matching to a specified UV completion. The couplings $c_{W1}$ and $c_{W2}$ are the leading contributions to spin-dependent scattering at low-velocity.

\begin{figure}[t]\centering
\includegraphics[scale=0.4]{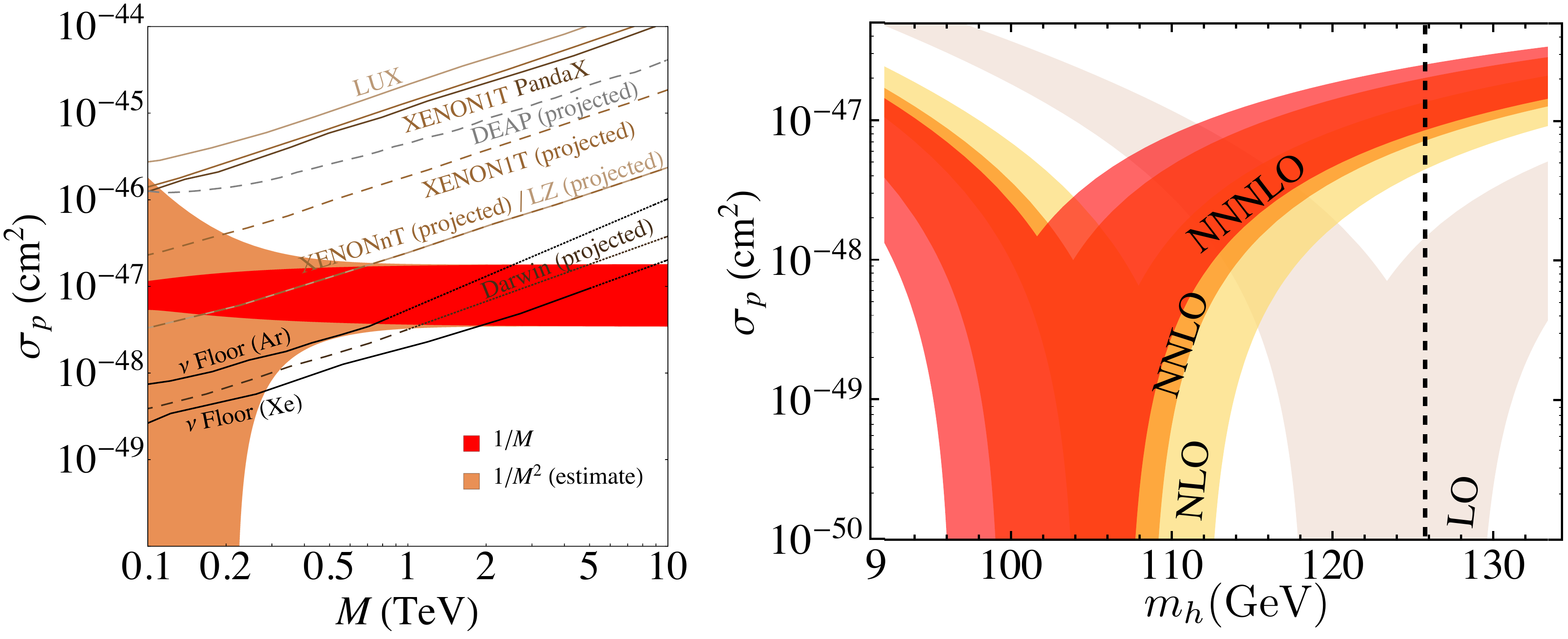}
\caption{(Left) WIMP-proton scattering cross section as a function of the WIMP mass for a Majorana triplet. For $M \gtrsim 700 {\rm GeV}$, the prediction is given by the universal $M \to \infty$ limit. The red band represents perturbative and hadronic input uncertainties, and includes sample model-dependent $1/M$ corrections. The orange band estimates the impact of $1/M^2$ corrections. Figure taken from Ref.~\cite{Chen:2018uqz}.  (Right) The impact of perturbative QCD corrections on the universal cross section is illustrated as a function of the Higgs mass. Bands represent predictions including higher-order contributions (as labeled) to the running and matching calculations below the weak-scale. Figure taken from Ref.~\cite{Hill:2013hoa}.}
\label{fig:heavy_direct}
\end{figure}

The simple universal heavy WIMP limit, obtained by taking $M \to \infty$ in Eq.~\ref{eq:HWlagrangian}, was analyzed in Refs.~\cite{Hill:2011be,Hill:2013hoa}, and revealed that generic amplitude-level cancellations suppress the low-velocity WIMP-nucleon cross section by orders of magnitude compared to simple estimates. For instance, the cross section for the triplet was found to be $\sim 10^{-47}\,{\rm cm}^2$, and even smaller for the doublet. See Fig.~\ref{fig:heavy_direct}. In the presence of such cancellations, formally subleading effects can become relevant. The impact of corrections from perturbative QCD is illustrated in the right panel of Fig.~\ref{fig:heavy_direct}. The effect of the $c_H$ parameter was explored in Refs.~\cite{Chen:2018uqz,Chen:2019gtm}, while the effect of multinucleon matrix elements was studied in Ref.~\cite{Chen:2019gtm}.  

While these small cross sections make heavy WIMPs more challenging to probe in direct detection experiments, the model-independence of the heavy WIMP limit provides a benchmark that can be precisely constrained by studying the underlying Standard Model interactions. This includes contributions from matching at the weak scale~\cite{Hill:2014yka}, from running between and matching at quark thresholds, and from evaluating hadronic matrix elements~\cite{Hill:2014yxa}. These calculations also involve a number of effective field theory methods such as those used in perturbative QCD and chiral perturbation theory. See Sec.~\ref{sec:direct:detection} for further discussion.

Heavy particle methods have also been applied for studying the direct detection phenomenology for the case of an additional singlet (bino)~\cite{Berlin:2015njh}, and have been extended and applied for resummation of large logarithms $\sim \log M /m_W$ relevant for the case of indirect detection, as discussed in Sec.~\ref{sec:inddet}, as well as for analyzing corrections to Sommerfeld enhancement, as discussed in Sec.~\ref{sec:somm}.

\section{Heavy DM: Indirect Detection}
\label{sec:inddet}

In many dark matter scenarios, we can observe a flux of its annihilation or decay products from the ambient dark matter concentrations throughout the Universe. (See Ref.~\cite{Slatyer:2021qgc} for a recent overview of this ``indirect detection of dark matter.'') As in Sec.~\ref{sec:HeavyDM}, getting an observable signal is straightforwardly realized by augmenting the SM with additional particles charged under the electroweak $SU(2)_W \times U(1)_Y$.  However, there are other possibilities, such as a coupling $\mathcal{L} \supset \bar \chi \chi \, \bar \psi \psi,$ where $\chi$ is the WIMP DM field that may be a SM singlet, and $\psi$ is some SM field.  The case of $\psi = H$ is the so-called ``Higgs portal,'' and if $\chi$ is itself a scalar, then the interaction even arises from a renormalizable operator \cite{Silveira:1985rk} ({\it cf.}~\cite{Arcadi:2019lka} for a current review).

In indirect detection, the presence of WIMP-WIMP interactions gives a richer structure to the heavy WIMP effective theory.  If the DM is charged under the electroweak force (or another force with light mediators), then it is subject to a long-range potential.  This can boost its annihilation rate via Sommerfeld enhancement, which is further detailed in Sec.~\ref{sec:somm}.  Additionally, the DM can bind into short-lived wimponium states, and for certain scenarios like the $SU(2)_W$ quintuplet, the capture photons are nearly detectable with present experiments for certain masses \cite{Mitridate:2017izz}.  Furthermore, the quintuplet offers an example of $\mathcal{O}(10\%)$ modification to the annihilation rate to energetic photons from the additional wimponium channels~\cite{Baumgart:2022yyy}.

For sufficiently heavy dark matter ($M \gtrsim 1$ TeV) one must account for the presence of energies parametrically above the weak scale.  Even in this regime though, the electroweak symmetry remains broken.  Thus, even observables like the semi-inclusive annihilation rate to $\gamma + X$ exhibit double-log-enhanced radiative corrections from Bloch-Nordsieck violation \cite{Ciafaloni:1998xg,Ciafaloni:1999ub,Ciafaloni:2000df}.  In the $M_\chi \gg m_W$ regime, the event kinematics relevant for indirect detection resemble those of a collider experiment, but with showers of electroweak (rather than QCD) radiation accompanying the photon and recoil ``jets.''  The appropriate EFT to sum the large logs of the resulting scale hierarchies is soft-collinear effective theory (SCET) \cite{Bauer:2000ew,Bauer:2000yr,Bauer:2001ct,Bauer:2001yt}.  This was initially developed for WIMP annihilation in a series of papers studying $SU(2)_W$ triplet or ``wino'' dark matter \cite{Baumgart:2014vma,Bauer:2014ula,Ovanesyan:2014fwa,Baumgart:2014saa,Baumgart:2017nsr,Baumgart:2018yed}, but one can go to other electroweak multiplets like higgsino \cite{Baumgart:2015bpa,Beneke:2019gtg} or quintuplet \cite{Baumgart:2022yyy} with similar machinery since many of the operator structures are the same.  The main ingredients are the soft Wilson lines and collinear gauge fields whose interactions are highly constrained by the large gauge symmetry of SCET.  The latter is expressed as
\beq
\mathcal{B}_{n\perp}^\mu(x)
= \frac{1}{g}\Bigl  [W_{n}^\dagger(x)\,i  D_{{n}\perp}^\mu W_{n}(x)\Bigr],
\eeq
where $W_{n}(x)$ is a collinear Wilson line given by 
\beq
W_n(x) =\left[ \, \sum\limits_{\text{perms}} \exp \left(  -\frac{g}{\bar{n}\cdot \mathcal{P}}\, \bar n \cdot A_n(x)  \right) \right]\!.
\eeq
See {\it e.g.}~Ref.~\cite{Baumgart:2017nsr} for a careful definition of these expressions. It is straightforward to include SCET fields for SM fermions and the Higgs, as well.

Simple electroweak WIMP scenarios yield multi-TeV photon line signals, making them natural targets for gamma-ray telescopes.  Nonetheless, a benefit of developing the electroweak SCET is that it allows (along with standard QCD SCET) a first-principles resummation of the final state shower of radiation.  Thus, the modification for different experimental scenarios is straightforward, as the same EFT can provide results for continuum photon, $e^+/e^-$, neutrino, or cosmic ray final states.  Changing the initial dark matter process is simple, too, as one just requires a different high-scale annihilation or decay operator (like that of the Higgs portal) before passing to the SM SCET.

\begin{figure}[t]\centering
\includegraphics[scale=0.35]{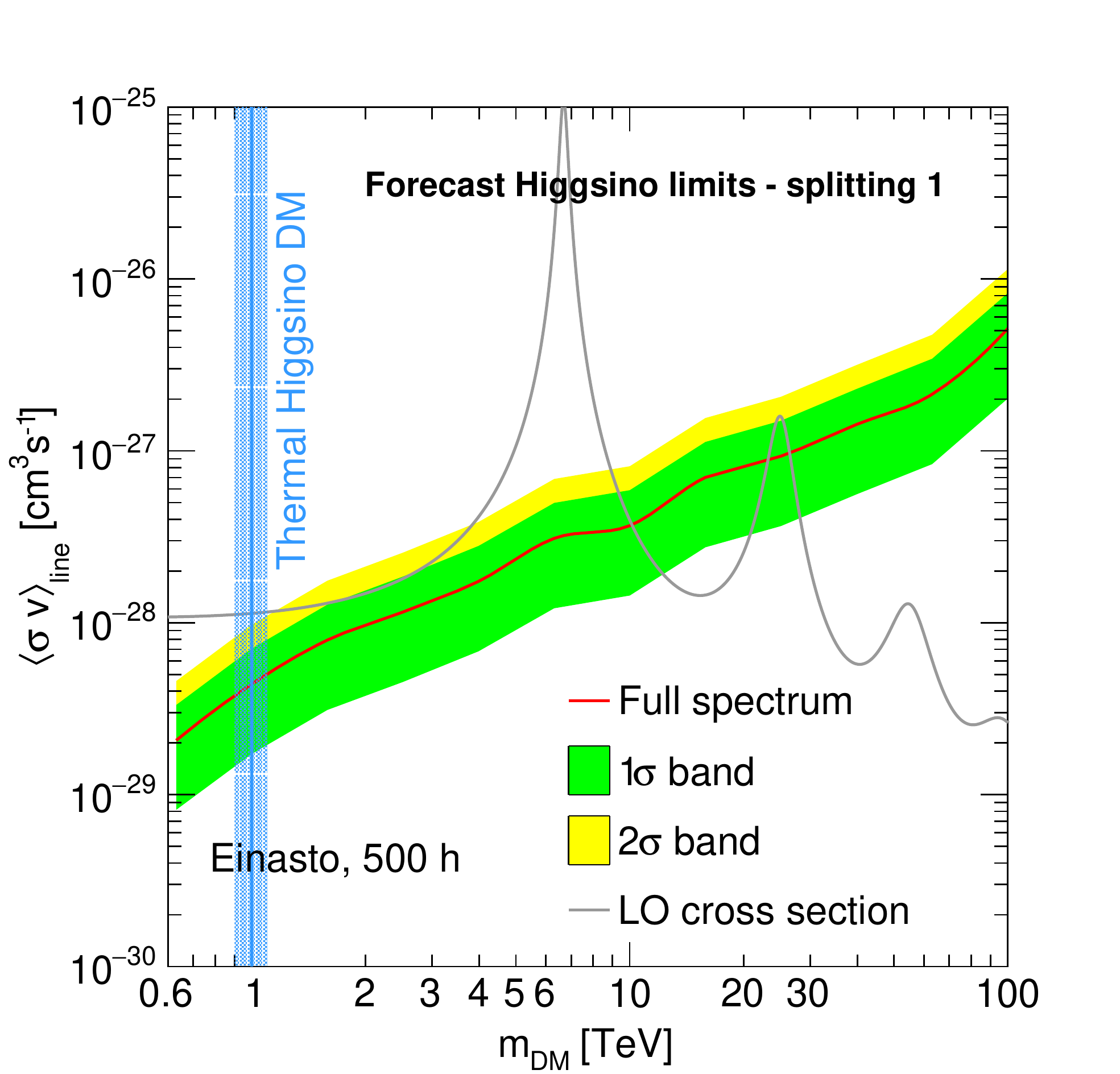}\hspace{0.5cm}
\includegraphics[scale=0.35]{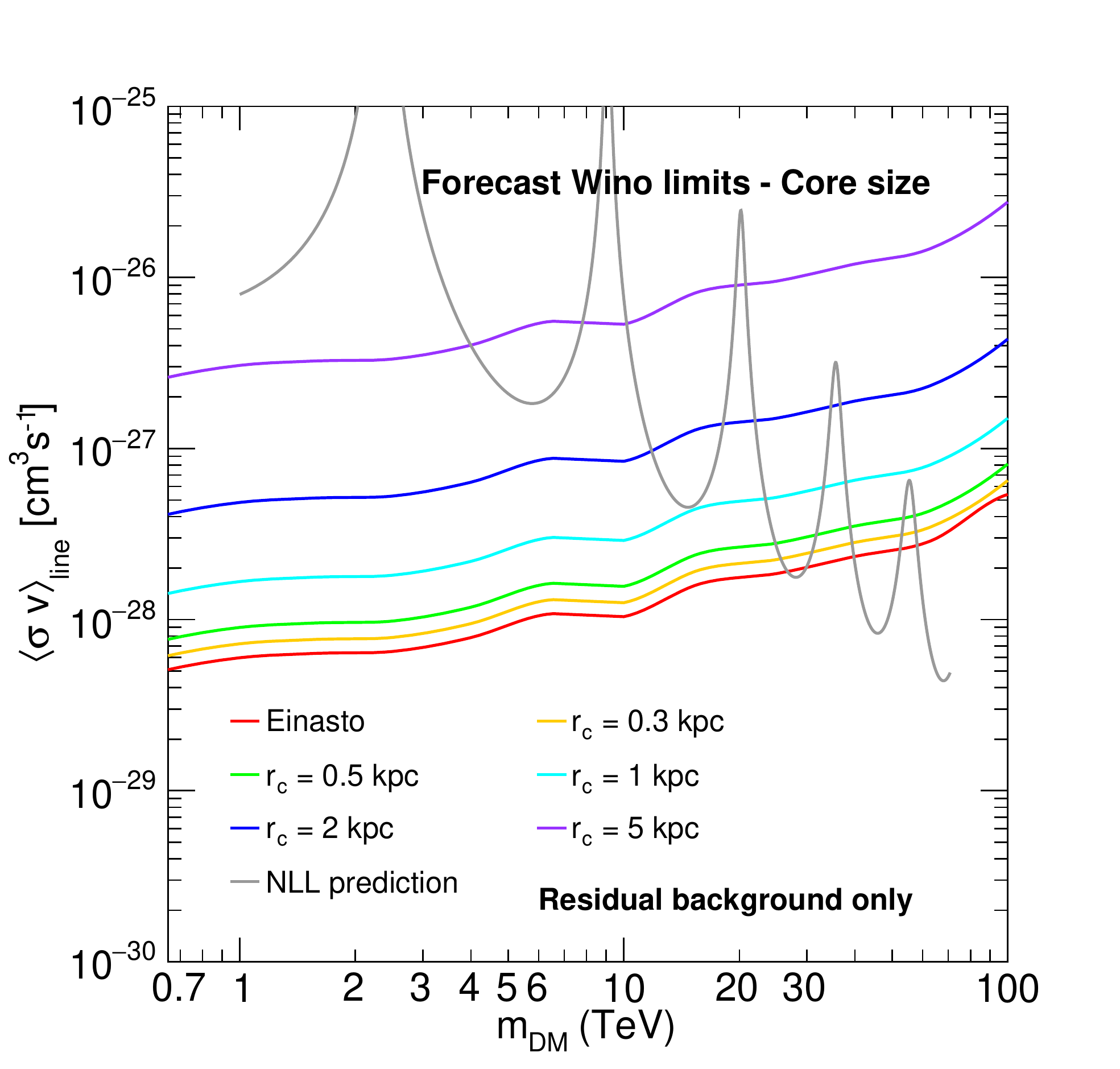}
\caption{Results reproduced from Ref.~\cite{Rinchiuso:2020skh} showing the sensitivity of CTA to Higgsino (left) and wino (right) DM annihilation in the galactic center.
For the Higgsino, the result suggest that CTA could be sensitive to the cross-section predicted at the thermal mass $\sim$1\,TeV.
However, the theory prediction does not include the effects of the EFT of Heavy DM, which could easily shift the result in either direction by a factor of few, and provides a clear motivation for this work.
For the wino, one sees that even with extremely conservative assumptions about the DM content of the inner galaxy, corresponding to a very large core size, the predicted cross section at the thermal mass of $\sim$3\,TeV will be excluded ({\it cf.} at present this is not excluded by HESS~\cite{Rinchiuso:2018ajn}).
We refer to Ref.~\cite{Rinchiuso:2020skh} for additional details.}
\label{fig:higgsino-CTA}
\end{figure}

As stated above, a primary application of the heavy DM EFT framework is to indirect DM searches performed with terrestrial imaging atmospheric Cherenkov telescopes (IACTs).
When a high-energy gamma-ray impacts the atmosphere, the charged particles present in the resulting shower produce Cherenkov radiation that can be detected on the Earth's surface on sufficiently dark nights.
Detecting this radiation, an IACT can reconstruct the incident photon with considerable angular and energy resolution.
In exactly this manner, an IACT can search for the photons generated by heavy DM annihilating in the galactic center.
The current generation of instruments searching for DM include MAGIC~\cite{FlixMolina:2005hv,MAGIC:2016xys}, Veritas~\cite{Weekes:2001pd,VERITAS:2006lyc,Geringer-Sameth:2013cxy}, and HESS~\cite{Hinton:2004eu,HESS:2013rld,Rinchiuso:2018ajn,Hryczuk:2019nql}.
Observations of the center of the Milky Way -- where the brightest signal from DM annihilation is expected -- already place the thermal wino under tension.
As explored in Ref.~\cite{Rinchiuso:2018ajn}, to avoid these constraints one can exploit our uncertainty in the DM density in the inner part of the Galaxy.
For instance, if instead of steadily increasing towards the galactic center the DM is cored within $\sim$2\,kpc, then the predicted signal flux can be reduced below the current limits.
The next generation IACT, the Cherenkov Telescope Array (CTA)~\cite{CTAConsortium:2010umy,Silverwood:2014yza,Lefranc:2015pza,Hryczuk:2019nql}, will be able to remove this caveat~\cite{Rinchiuso:2020skh} ({\it cf.}~Fig.~\ref{fig:higgsino-CTA}).
As shown in Ref.~\cite{Rinchiuso:2020skh}, CTA may also be able to probe the long sought after thermal Higgsino, a plot from that work demonstrating this is reproduced in Fig.~\ref{fig:higgsino-CTA}.
Yet this result did not draw on the EFT of Heavy DM, and whether this enhances or reduces the annihilation cross-section will determine the fate of this DM candidate in the coming years.

An interesting feature of the SCET used for indirect detection is the presence of two types of scale hierarchies.  We straightforwardly have large logs of the form $\log(M_\chi/m_W)$.    
A key aspect of IACTs that further influences the required EFT calculations is the energy resolution.
Thus, one also picks up large logs of the form $\log(1-z_\text{cut})$, where $z_\text{cut}$ is the energy fraction of the WIMP mass below which a photon cannot have come from a simple 2 $\rightarrow$ 2 annihilation.
For ${\cal O}(\text{TeV})$ energy gamma-rays, existing instruments such as HESS can achieve $\Delta E/E \sim 10\%$, a value CTA will improve to $\sim 5\%$ in the future.
In either event, the resolution is insufficient to distinguish between the line spectrum -- photons emerging from two-body final states, such as $\gamma \gamma$ and $\gamma Z$ -- and those originating from so-called endpoint photons, which cannot be distinguished from the line photons given the finite resolution.
This is why the inclusion of endpoint photons is critical, and can impact the IACT sensitivities by an ${\cal O}(1)$ factor~\cite{Baumgart:2017nsr,Baumgart:2018yed}.
Of course it is also important to include photon contributions which can be distinguished from the line spectrum, broadly categorised as continuum photons.
For example, these can emerge from the decay products of a $W^+ W^-$ final state, and can be incorporated using results provided in Refs.~\cite{Cirelli:2010xx,Bauer:2020jay}.
We note that continuum photons play a key role in CTA's sensitivity to the thermal Higgsino~\cite{Rinchiuso:2020skh}, as the telescope will have significantly enhanced sensitivity to lower energy gamma-rays than previous IACTs.
In \cite{Beneke:2018ssm,Beneke:2019vhz,Beneke:2019gtg,Beneke:2022eci}, the authors used the numerical coincidence of $(1-z_\text{cut})$ and $m_W/M_\chi$ over much of the range of current and upcoming experiments for a simple EFT treatment of the wino and Higgsino.  Equating these scale hierarchies is not possible to do though, for the full dataset anticipated for the CTA experiment~\cite{Rinchiuso:2020skh}.  Thus, the Higgsino remains a compelling target for further theoretical study.

\section{Heavy DM: Corrections to Sommerfeld Enhancement}
\label{sec:somm}

WIMP dark matter exhibits exciting phenomenology. As noted by Hisano et al. \cite{Hisano:2003ec,Hisano:2004ds,Hisano:2006nn}, the annihilation cross-section is substantially altered when the DM mass is above about 1~TeV due to the so-called Sommerfeld effect.  
In the non-relativistic regime, as the relative DM velocity approaches zero, $v \to 0$, the ladder diagrams formed by exchanges of weak gauge bosons between the DM particles are responsible for the enhanced corrections. Each loop is suppressed by a weak coupling constant $\alpha_2$, but it receives an enhancement by a factor $M_\chi/m_W$, where $M_\chi$ is the DM mass. When $\alpha_2  \times M_\chi/m_W \sim 1$, the ladder diagrams must be summed to all orders in perturbation theory. 
This section describes an EFT formalism allowing us to compute DM annihilation cross-sections and relic density for models exhibiting Sommerfeld enhancement. The main advantage offered by the EFT formalism is systematic expansion and easiness of including higher-order corrections. 

Potential non-relativistic EFT (PNRQED) is a natural framework for computing Sommerfeld factors \cite{Beneke:2012tg,Hellmann:2013jxa,Beneke:2014gja}. PNRQED has been initially developed in the contexts of QED and QCD bound states and threshold problems \cite{Pineda:1997bj,Beneke:1998jj,Brambilla:1999xf}. It is obtained form the full theory after integrating out off-shell degrees of freedom. The counting parameter is given by velocity $v\ll 1$. 
The Lagrangian, written in terms of potential DM fields $\chi_{i}$, which form a multiplet under $SU(2)_W$,  relevant for computations of annihilation cross section, is given by
\begin{align}
\label{eq:LPNRDM}
\mathcal{L}_{\rm PNRDM} &= \sum_{i} \chi_{i}^\dagger(x) \left(i
D^0_i(t,\mathbf{0}) - \delta m_i+ \frac{\boldsymbol{\partial}^2}{2M_\chi} \right)\chi_{i}(x) \nonumber\\ 
                        &-\,\sum_{\{i,j\},\{k,l\}} \int d^3\mathbf{r}\,
V_{(ij),(kl)}(r)\, \chi_{k}^\dagger(t,\mathbf{x})
\chi_{l}^\dagger(t,\mathbf{x}+\mathbf{r}) \chi_{i}(t,\mathbf{x})
\chi_{j}(t,\mathbf{x}+\mathbf{r}) \, .
\end{align}
The mass splitting between the multiplet members, which has a substantial impact on phenomenology \cite{Slatyer:2009vg}, is denoted by $\delta m_i$ and $V_{(ij),(kl)}(r)$ are the potentials. At the leading order,  $V_{(ij),(kl)}(r)$ is a combination of Coulomb terms due to photon exchange and Yukawa potentials arising from massive gauge bosons.
For example, for DM forming a Majorana $SU(2)_W$ triplet, we decompose the two-particle states into states with definite total angular momentum and spin, and find that the potential for the charge-neutral sector for the $S$ wave spin-singlet $^1S_0$ and triplet $^3S_1$ configurations reads
\begin{align}
  V^{Q=0}(r)({}^1S_0) &= \begin{pmatrix} 0 & & -\sqrt{2} \alpha_2 \frac{e^{-m_W
    r}}{r} \\
        - \sqrt{2} \alpha_2 \frac{e^{-m_W r}}{r} & & -\frac{\alpha}{r} -
        \alpha_2 c_W^2 \frac{e^{-m_Z r}}{r} 
    \end{pmatrix}\,, \\[0.2cm]
    V^{Q=0}(r)({}^3S_1) &= \begin{pmatrix} 0 & & 0 \\
        0 & & -\frac{\alpha}{r} - \alpha_2 c_W^2 \frac{e^{-m_Z r}}{r} 
    \end{pmatrix} \, .
\end{align}

The precise predictions for DM annihilation rates and relic abundance require inclusion of the next-to-leading order (NLO)  potentials \cite{Beneke:2019qaa,Beneke:2020vff,Urban:2021cdu}, especially near the location of the resonance. These corrections come from integrating out regions with soft momentum scaling. 
Often, the corrections affect the asymptotic behavior of the potentials. For example, massless fermions generate long-distance contributions, which dominate the behavior for large $r$. In the $\chi^0 \chi^0 \to \chi^+ \chi^-$ channel, at the LO, only short-range Yukawa-type potential appears, but the NLO corrections change the asymptotic behavior to power-like $\delta V(r)\sim 1/r^5 $ behavior.  

\begin{figure}
    \centering
    \includegraphics[width=0.49\textwidth]{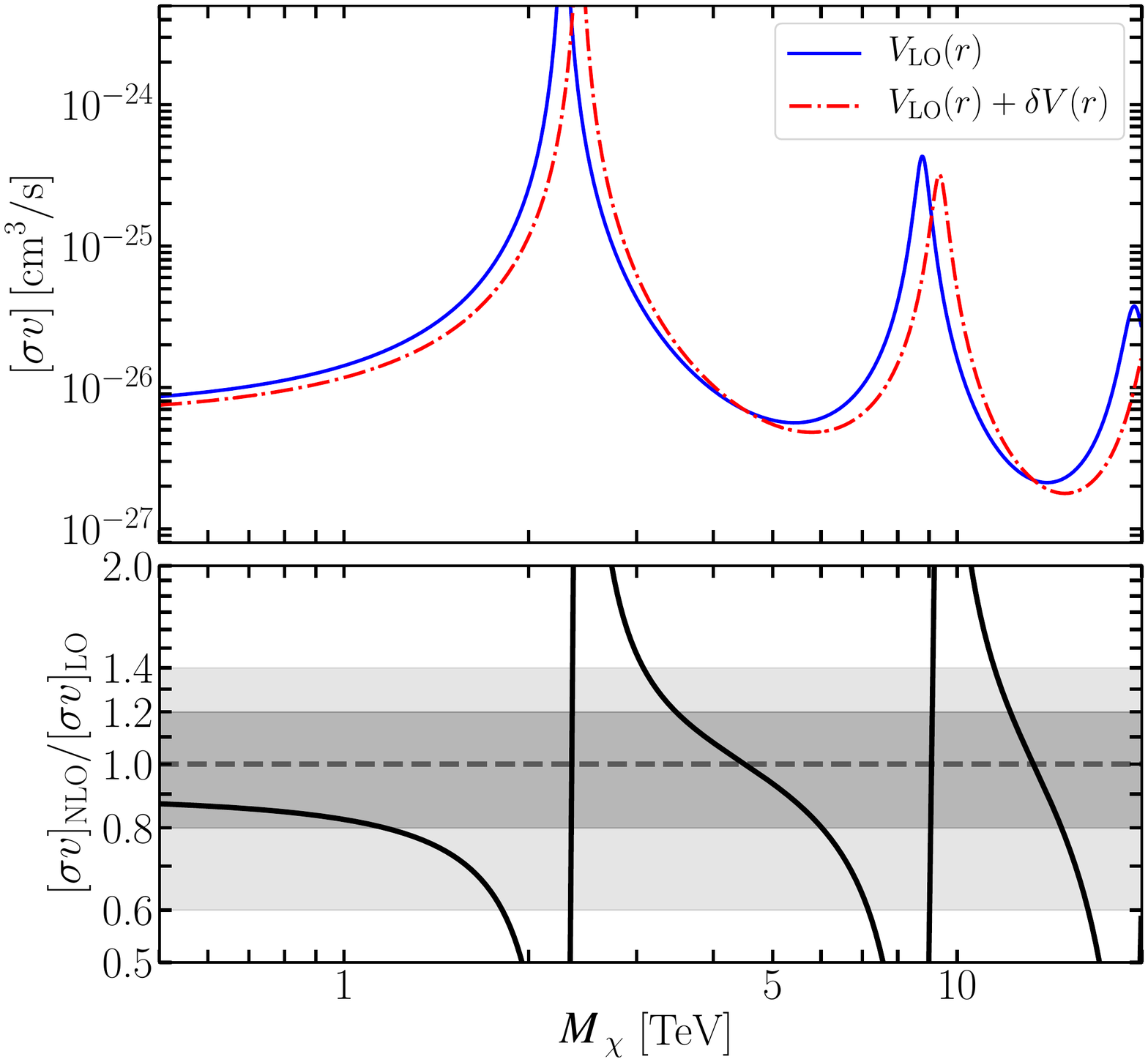}
    \includegraphics[width=0.49\textwidth]{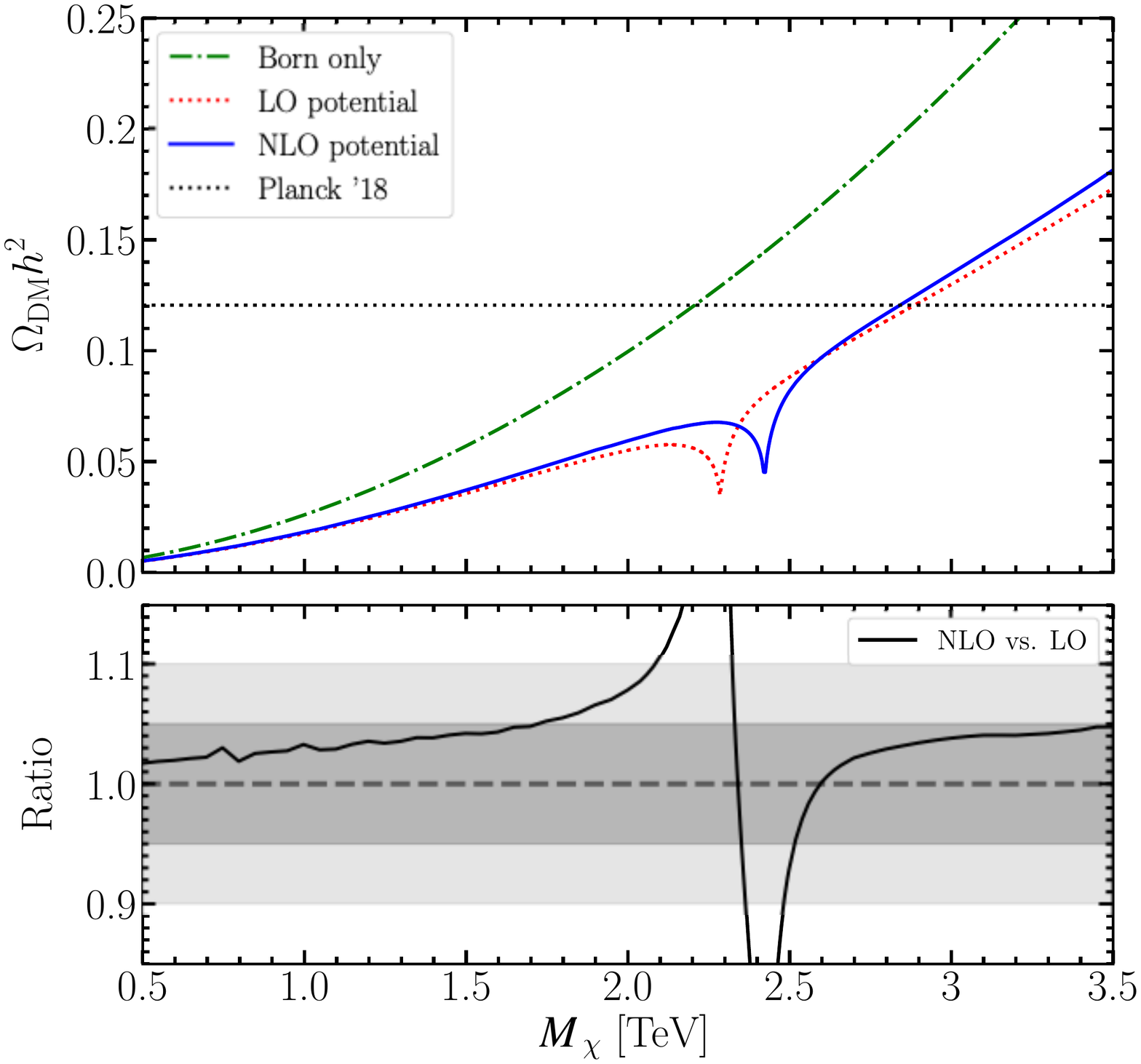}
    
    \caption{On the left: The annihilation rate $\chi^0\chi^0 \to \gamma+X$, computed with the LO (blue) and the NLO (red) potential as a function of DM mass. The ratio NLO/LO is shown in the lower panel with gray bands indicating the range where the correction stays below 20\% and 40\%. On the right: The relic abundance as a function of DM mass computed with Born cross sections  (green), including LO (red) and NLO (blue) Sommerfeld enhancement. The horizontal line corresponds to the observed relic abundance. The lower panel illustrates the ratio of the NLO to LO Sommerfeld-corrected cross section. Plots reproduced from \cite{Beneke:2019qaa,Beneke:2020vff}.}
    \label{fig:NLOpotential}
\end{figure}

The $\mathcal{L}_{\rm PNRDM}$ describes the dynamics of the heavy particles. To describe annihilation, which is a short-distance process, it must be supplemented by higher-dimensional operators $\delta \mathcal L_\text{ann}$ \cite{Beneke:2012tg,Hellmann:2013jxa,Beneke:2014gja}, whose Wilson coefficient are determined by matching the four-fermion operators on the full theory amplitude for $\chi_i \chi_j \to \chi_l \chi_k$.

The next step is to compute the co-annihilation cross-sections as functions of relative velocity by taking relevant matrix elements of $\delta \mathcal L_\text{ann}$. The resulting expression combines Wilson coefficients and Sommerfeld factors for individual channels. 
The Sommerfeld factors are evaluated by solving the Schr\"odinger equation and are given by the incoming DM particles' scattering wave-functions evaluated at the relative distance equal to zero.
In practice, the solution is performed numerically, as the analytical solutions of the Schr\"odinger equation with Yukawa potential are not known.

From here, determining the DM abundance requires the computation of the thermally averaged cross-section $\langle \sigma_{\rm eff} v \rangle$ for all possible co-annihilation channels.  
To compute the thermal relic DM abundance we solve the Boltzmann equation
\begin{align}
    \frac{d n(t)}{dt} + 3 H(t) n(t)  = - \langle \sigma_{\rm eff} v \rangle (n^2(t)-n^2_{\rm eq } (t) ),
\end{align}
where $n(t)$ is the total number density, $H(t)$ is the Hubble rate and the   $n_{\rm eq}(t)$ is the total equilibrium number density.  Computing the present-day yield, we can obtain the relic density. 
The measured value of the DM abundance $\Omega_{\rm DM} h^2 =  0.1205$ is reproduced in a simple Wino model for $M_\chi =2.886\, \rm TeV$ using LO potential and  $M_\chi = 2.842\, \rm TeV$ for NLO potential, see the right plot in Fig.~\ref{fig:NLOpotential}. A $2\%$ shift is typical size for electroweak corrections away for the resonance region. The effect is more prominent for models where the location of the resonance plays a significant role. The resonance is achieved for  $M_\chi = 2.282\,\rm TeV$($M_\chi = 2.419\, \rm TeV$) for the LO (NLO) potential. 

As the DM velocity is now much smaller than during the freeze-out, the NLO corrections are more pertinent for the present-day annihilation rate, shown in the left plot in Fig.~\ref{fig:NLOpotential}, as a function of DM mass. The NLO corrections exceed $20\%$ for a wide range of phenomenologically viable DM masses. Therefore, it is of paramount importance that NLO potential is included on a par with the resummation of soft and collinear effects.

The described non-relativistic EFT formalism can be applied to arbitrary models of heavy DM interacting with much lighter bosons. The NLO corrections to the non-relativistic potential can be found directly from \cite{Beneke:2020vff} for $SU(2)_W$.  They are also implemented in the code \texttt{DM$\gamma$Spec} \cite{Beneke:2022eci}, which allows computing DM annihilation spectra $\chi \chi \to \gamma +X$ that additionally include resummation of large electroweak corrections \cite{Ovanesyan:2014fwa,Ovanesyan:2016vkk,Baumgart:2017nsr,Baumgart:2018yed,Beneke:2018ssm,Beneke:2019gtg,Beneke:2019vhz}, see Sec.~\ref{sec:inddet}.

\section{Direct detection DM EFT for general mediators}
\label{sec:direct:detection}
 For a large class of DM models, the
physics of direct detection experiments, where DM scatters on nuclei, can be described using
effective field theories~\cite{Bagnasco:1993st,
  Pospelov:2000bq, Kurylov:2003ra, Kopp:2009qt, Fan:2010gt, Hill:2011be,
  Cirigliano:2012pq,Hill:2013hoa, Fitzpatrick:2012ix,
  Fitzpatrick:2012ib, Menendez:2012tm, Anand:2013yka, Klos:2013rwa,
  DelNobile:2013sia, Barello:2014uda, Hill:2014yxa, Catena:2014uqa,
  Hoferichter:2015ipa,Berlin:2015njh,
 Hoferichter:2016nvd, Bishara:2016hek,
  Bishara:2017pfq, DEramo:2016gos, Bishara:2017nnn, DEramo:2017zqw,
  Brod:2018ust, Brod:2017bsw,Chen:2018uqz,Hoferichter:2020osn,Hoferichter:2018acd,Aebischer:2022wnl,Brod:2021xbb}.  The reason is that the
momentum transfer $q$ for DM scattering on a nucleus is small, typically
less than 200\,MeV, so that the effect of forces mediated by particles heavier than this scale can be described by an EFT.
Furthermore, the interactions between the DM and the nucleus can be organized via a power counting parameter $q/\Lambda_\chi$ where $q$ is the momentum transfer and $\Lambda_\chi$ is the chiral symmetry breaking scale, so that the interactions are organized by their chiral dimension~\cite{Weinberg:1990rz,Weinberg:1991um}.
  
\begin{figure}[t]\centering
\includegraphics[scale=0.9]{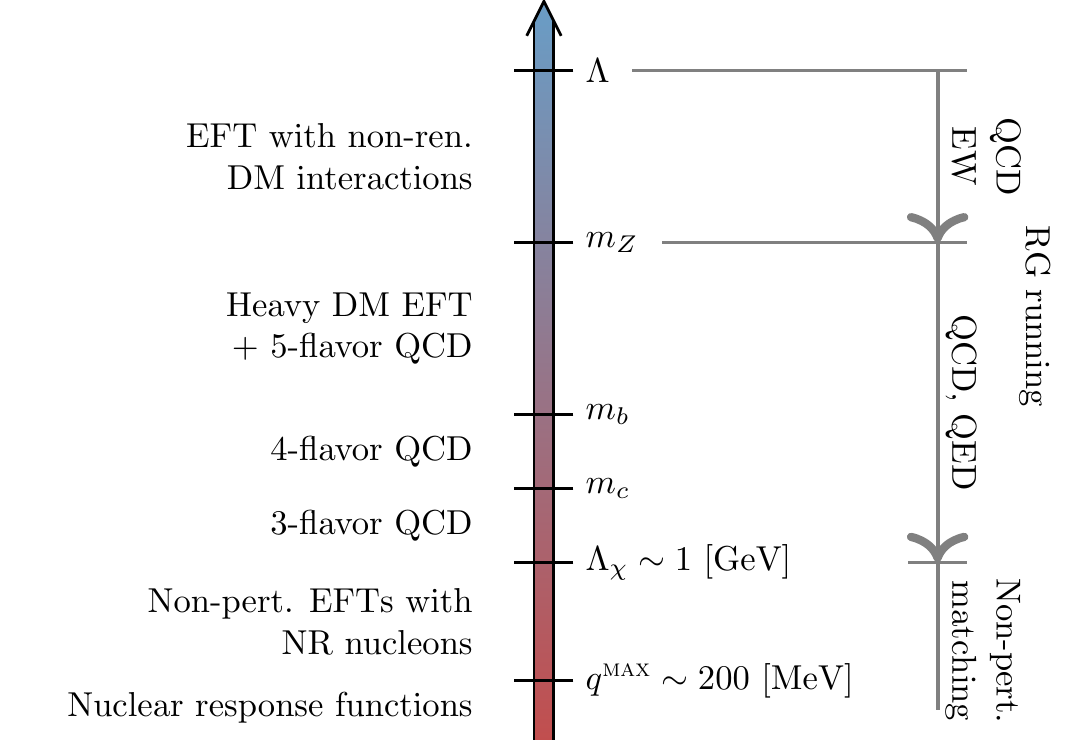}
\caption{The tower of EFTs linking the UV scale $\Lambda$ to the scale
  of interactions between the nucleons and the DM. Plot reproduced from \cite{Bishara:2018vix}.}
\label{fig:eft-tower}
\end{figure}

The construction of DM EFTs for DM scattering on nuclei has two goals. The first goal is to
compare the results of direct detection experiments that use different
target materials in a model-independent way. To achieve this, a DM EFT
valid at a scale $\mu\simeq 2$ GeV can be constructed. The second goal
is to connect the results of the direct detection experiments to the
physics at much higher scales: indirect DM searches, DM production at
the LHC, and ultimately to the full UV theory of DM. In this case one
can construct a tower of EFTs, see Fig.~\ref{fig:eft-tower}.
  
 At low
energies, the interactions of DM with the SM can be parametrized in two
different ways. The first is in terms of an EFT where DM interacts
with quarks, gluons, and photons (three-flavor DM EFT at $\mu\simeq 2$
GeV~\cite{Bishara:2017pfq}). The second option is the Galilean
invariant EFT, or NR~EFT, in which DM
interacts with non-relativistic neutrons and protons~\cite{Fitzpatrick:2012ib, Fitzpatrick:2012ix,
  Anand:2013yka}. 
Here, we focus first on the three-flavor DM EFT, while parametrization using NR~EFT is discussed in Sec.~\ref{sec:nucleon}.
 
 \subsection{Three flavor DM EFT}
In the three-flavor DM EFT, the operators are organized in terms of operator dimensions so that the effective Lagrangian takes the form
\beq
\label{eq:L:DMEFT}
{\cal L}_{\rm DM EFT}= \sum_{d,a} \frac{{\cal C}_a^{(d)}}{\Lambda^{d-4}} {\mathcal
  Q}_a^{(d)} ,
  \eeq
   where ${\cal C}_a^{(d)}$ are
dimensionless Wilson coefficients, and $\Lambda$ is the typical scale of
the UV theory for DM. The sum is over different operators, ${\mathcal{Q}_a}$, of
dimension $d$. An example
of a $d=6$ operator for fermionic DM is $(\bar \chi \gamma^\mu \chi)(\bar
q\gamma_\mu q)$ for a vectorial interaction; a typical $d=7$
operator is $(\bar \chi \chi)G_{\mu \nu}^a G^{a \mu \nu}$ for a scalar interaction.
The full basis of operators for scalar and fermion DM of up to and including dimension seven can be
found in~\cite{Brod:2017bsw}, while the basis for vector DM was given in \cite{Aebischer:2022wnl}. The expansion in \eqref{eq:L:DMEFT} assumes that there is a mass gap between the DM candidate and the mediators,  $m_\chi \ll \Lambda$. If that is not the case, i.e., if the mediators have a mass comparable to that of the DM or are even lighter, the appropriate description is in terms of the Heavy DM EFT discussed in Sec.~\ref{sec:HeavyDM}.

The DM EFT (or Heavy DM EFT) facilitates a model-independent
comparison of direct detection experiments, which is best done as
follows: the Wilson coefficients ${\cal C}_a^{(d)}$ in three-flavor DM
EFT can be treated as unknown and are freely varied in a fit. The DM
scattering rates for different targets are obtained by first matching
onto NR~EFT, or more generally the chiral
EFT~\cite{Hoferichter:2015ipa, Bishara:2016hek,
  Cirigliano:2013zta}. The leading-order expressions are given in
Ref.~\cite{Bishara:2017pfq}, but sub-leading contributions, such as
the effect of two-body currents, can also be
included~\cite{Hoferichter:2018acd, Hoferichter:2019uwa,
  Cirigliano:2013zta}.  They are then used to obtain the DM--nucleus
scattering cross sections. For both steps one can use public codes,
{\tt DirectDM} \cite{Bishara:2017nnn} in combination with {\tt
  DMFormFactor} \cite{Anand:2013yka}, or {\tt ChiralEFT4DM}
\cite{Hoferichter:2018acd}.  For a general Dirac fermion DM, 25
independent combinations of Wilson coefficients ${\cal C}_a^{(d)}$
would be varied in three-flavor EFT fits when considering operators up to dimension-seven (excluding $d=7$ operators with derivatives) and at
leading order in the chiral expansion. While this is still a rather
large set of parameters, the benefit is that this approach captures
all UV models for DM of a given spin with mediators heavier
than a few hundred MeV.

One could instead entertain comparisons of direct detection
experiments by only allowing for a subset of Wilson coefficients to be
non-zero. The two well-known limits already widely used are
spin-independent and spin-dependent scattering, but one can extend
this to other well motivated benchmark choices, some of which we list
below:

\paragraph{Spin-independent (SI) scattering.} The operators $(\bar \chi \gamma^\mu \chi)(\bar q\gamma_\mu q)$, $(\bar \chi \chi)(\bar q q)$, $(\bar \chi \chi)(GG)$ all lead to spin-independent scattering. That is, at leading order in chiral expansion they give rise to NR~EFT operators, $ L_{\rm NR} = c_1^N \mathbb{1}_\chi \otimes \mathbb{1}_N$, where
$\mathbb{1}_{\chi} (\mathbb{1}_{N})$ are the number operators for DM
(nucleons) and the sum over $N=p,n$ is implied. For the comparison of direct detection experiments under
the assumption of SI scattering it suffices, therefore, to vary the SI
couplings of DM to protons and neutrons, $c_1^{p,n}$.

\paragraph{Spin-dependent (SD) scattering.} In the NR limit there are two types of interactions between the DM spin, $\vec S_\chi$, and nucleon spin, $\vec S_N$, $L_{\rm NR}=c_4^N \vec S\cdot \vec S_N+ c_6^N (\vec S_\chi \cdot {\vec q})\, (\vec S_N\cdot {\vec q})/m_N^2$. Conventionally, in the comparisons of direct detection experiments under the assumption of pure SD scattering, the $c_6^N$ are set to zero, which corresponds to taking just the tensor operator $\big(\bar \chi \sigma_{\mu\nu}\chi)(\bar q \sigma^{\mu\nu} q)$ in the three-flavor DM EFT to be nonzero. An alternative limit is to assume that only the axial--axial operator $\big(\bar \chi \gamma_\mu \gamma_5 \chi)(\bar q \gamma^\mu \gamma_5 q)$ is relevant, in which case both $c_4^N$ and $c_6^N$ are generated, the latter with a parametric size $c_6^N\sim m_N^2/(m_\pi^2+\vec{q}^2)$, but numerically only important for scattering of DM on heavier nuclei.

\begin{figure}[t]\centering
	\includegraphics[width=0.49\textwidth]{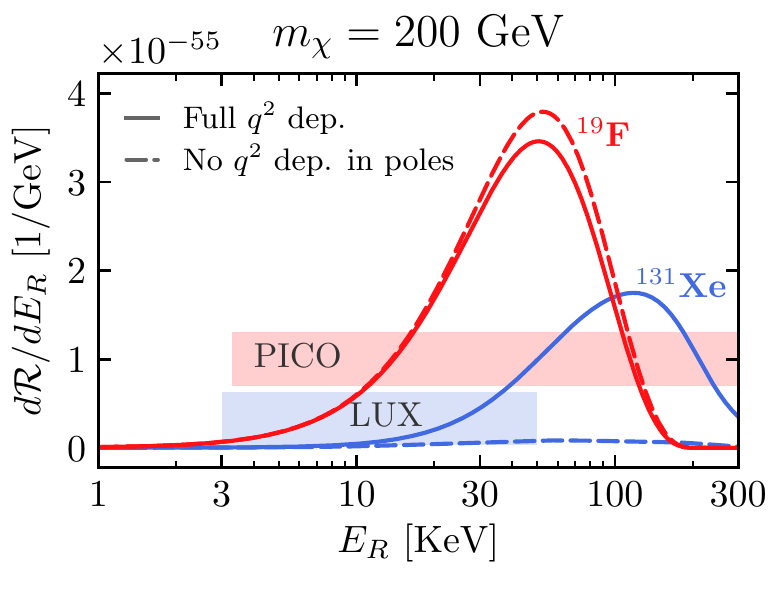}
	\includegraphics[width=0.49\textwidth]{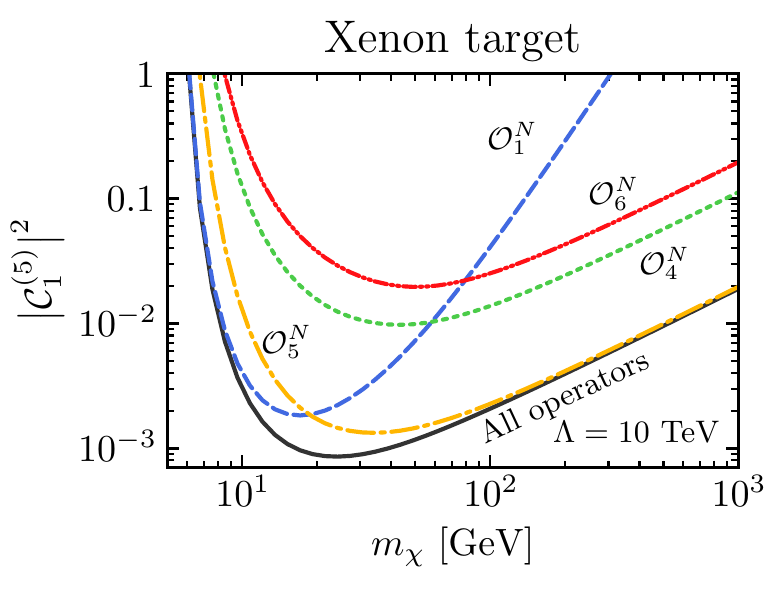}
	\caption{\label{fig:P} \emph{Left panel:} recoil energy dependence for the 	`pseudoscalar mediator' benchmark. The dashed lines show the artificial limit of neglecting terms proportional to $\vec q^2$.
	The shaded bands show the sensitivity windows for PICO and LUX.
	\emph{Right panel:} contributions to the exclusion curve from the various NR operators in the case of magnetic dipole moment DM scattering on xenon. 	Plots reproduced from \cite{Bishara:2018vix}.
}
\end{figure}

\paragraph{Gluophilic pseudoscalar mediator.} A  pseudoscalar mediator such as an axion-like particle (ALP) that couples to a fermionic DM and to the SM through the QCD anomaly (for simple realizations in 2HDM + singlet scenarios see, {\it e.g.},~\cite{LHCDarkMatterWorkingGroup:2018ufk}) would give rise to the three flavor operator $\bar{\chi} i \gamma_{5} \chi G^{a \mu \nu} \widetilde{G}_{\mu \nu}^{a}$. This results in a SD scattering with rather complicated $\vec q^2$ pole structure in the $c_6^N$ coefficient, after matching onto NR~EFT. The scattering rates on light nuclei, such as fluorine, and heavier nuclei, such as Xenon, can then be quite different from the conventional SD scattering benchmark discussed above, see Fig. \ref{fig:P} (left).

\paragraph*{Magnetic dipole DM.} If DM mediators are heavy and charged under the electroweak SM gauge group, but do not directly couple to quarks, the leading interaction between fermionic Dark Matter and the SM will be via the DM magnetic dipole moment, $\mathcal{Q}_1^{(5)}=\frac{e}{8\pi^2}\,\bar{\chi}\sigma^{\mu\nu}\chi F_{\mu\nu}$. This induces $c_1^N, c_4^N, c_5^N, c_6^N$ NR~EFT coefficients that give different relative contributions when scattering on heavy or light nuclear targets, and on whether DM is lighter or heavier. An example for xenon target is shown in the right panel of Fig. \ref{fig:P}.

\begin{figure}[t]\centering
	\includegraphics[width=0.49\textwidth]{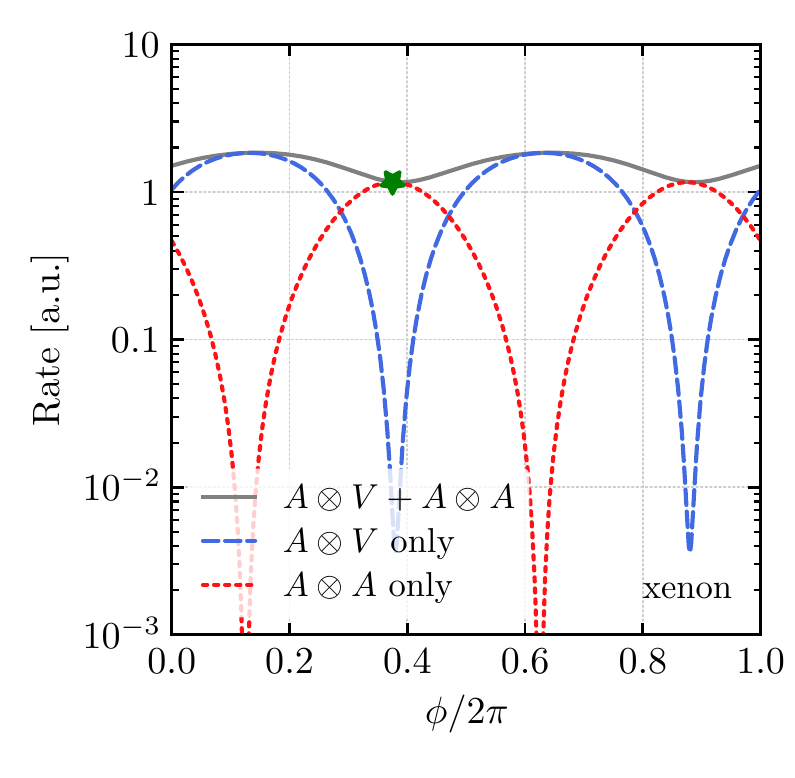}
	\includegraphics[width=0.49\textwidth]{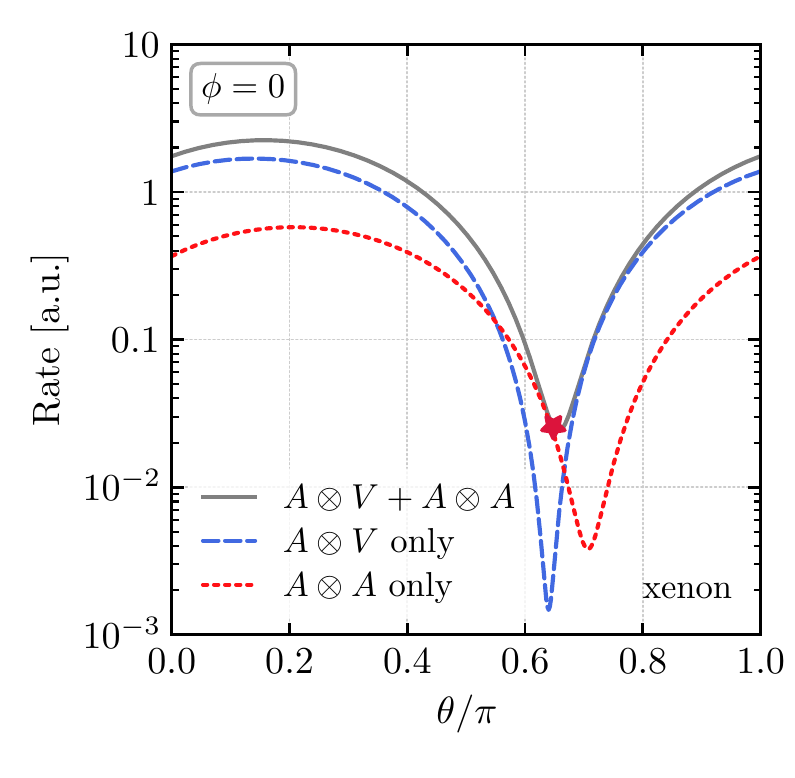}
	\caption{Relative event rate in direct detection on a Xenon target for $\tan\theta=2/(\sin\phi+\cos\phi)$ (left panel) and $\phi=\pi$ (right panel) for Majorana DM with a vector mediator.
	The stars denote potentially anomaly free models, which are studied in more detail in Ref.~\cite{Alanne:2022eem}. Plots reproduced from~\cite{Alanne:2022eem}.
	}
	\label{fig:isospin}
\end{figure}

\paragraph{Vector mediator for Majorana DM.} For Majorana DM there are four relevant operators in the three-flavor DM EFT: $(\bar \chi \gamma_\mu\gamma_5 \chi) (\bar q \gamma^\mu q)$,
$(\bar \chi\gamma_\mu\gamma_5 \chi)(\bar q \gamma^\mu \gamma_5 q)$, where $q=u,d$ (the $(\bar \chi \gamma_\mu \chi)$ currents vanish for Majorana fermions). The simplest UV realization for these a tree-level exchange of a heavy $Z'$, in which case electroweak gauge invariance requires that the four Wilson coefficients in three-flavor DM EFT are expressed in terms of just three operators, $(\bar\chi\gamma_\mu \gamma_5 \chi)(\bar Q_L \gamma^\mu Q_L)$, $(\bar\chi\gamma_\mu \gamma_5 \chi)(\bar u_R \gamma^\mu u_R)$, and $(\bar\chi\gamma_\mu \gamma_5 \chi)(\bar d_R \gamma^\mu d_R)$, with coefficients that can be parametrized as $({g^{\prime\,2}}/{\Lambda^2})\{ \cos\theta\,, \sin\theta\cos\phi\,,\sin\theta\sin\phi\} $, respectively~\cite{Alanne:2022eem}. 
This means that one cannot simultaneously suppress $A\otimes V$ and $A\otimes A$ operators for both up and down quarks. The effect is especially important for heavy nuclei, where $A\otimes V$ and $A\otimes A$ operators can give contributions of similar sizes to the direct-detection cross sections.  An example for scattering on xenon is shown in Fig.~\ref{fig:isospin}, with the two panels showing the dependence on parameters $\phi$ and $\theta$, when varying one and fixing the other.

 \subsection{EFT description of DM--nucleon interactions}
\label{sec:nucleon}

Instead of three flavor DM EFT, in which DM couples to quarks and gluons, one can also use NR-EFT \cite{Anand:2013yka,Fitzpatrick:2012ix,Fitzpatrick:2012ib}, in which DM couples to nonrelativistic nucleons, in order to parametrize the unknown interactions of DM. That is, instead of performing the series of matchings shown in Fig. \ref{fig:eft-tower}, starting with the three flavor DM EFT, and then nonperturbatively matching onto NR-EFT, followed by the calculation of nuclear response to DM scattering, one can just as easily use NR-EFT directly as the starting point, given that the DM interactions are not known yet. To leading order in the chiral expansion, DM only couples to single nucleon currents, and thus the effective DM interactions
take the form \cite{Anand:2013yka,Fitzpatrick:2012ix,Fitzpatrick:2012ib}
\beq
{\cal L}_{\rm NR}=\sum_a c_a^N(q) {\cal
  O}_a^N,
  \eeq
   where the interaction operators ${\cal O}_a^N$ involve the
non-relativistic DM and nucleons, with the latter only entering in the form of single nucleon currents. The UV physics is encoded in the coefficients $c_a^N$ where different UV models may match onto the same NR operators. For instance, both vector and scalar
mediators give rise to an operator $\mathbb{1}_\chi \otimes \mathbb{1}_N$, where
$\mathbb{1}_{\chi} (\mathbb{1}_{N})$ are the number operators for DM (nucleons). A single UV interaction may also result in several nonzero $c_a^N(q)$.

One main benefit of using NR-EFT is that the interactions are written trivially for any value of DM spin. Moreover, because short distance scales below the nucleon radius are integrated out, many different UV models can be subsumed into a smaller set of effective couplings; for instance, several operators in the three-flavor DM EFT all lead to the spin-independent operator in the NR EFT.  Furthermore, the nucleon-level interactions are more directly connected to nuclear response functions, which are determined by the internal structure of atomic nuclei. The nuclear response functions can be calculated once and for all, and thus also the cross sections for DM scattering on nuclei, as long as the coefficients $c_a^N(q)$ are given. The main drawback is that the cutoff of the NR-EFT is reduced compared to the three-flavor DM EFT, or more complete UV models, and therefore it has a smaller regime of validity; more over, the use of three flavor DM EFT makes the connection with the UV models of DM more immediate. The coefficients $c_a^N(q)$ are $q$ dependent, and in a strict EFT approach would be treated as a series expansion in powers of $q$ divided by the cutoff. However, light fields such as the photon and pion that are below the cutoff, and should in principle be included in the effective theory,  might nevertheless be `integrated out' for convenience, in which case they can generate additional poles in the $c_a^N(q)$ coefficients. 
Taking $c_a^N(q)$ to be constant in interpreting the data (see, {\it e.g.},~\cite{Xia:2018qgs, Liu:2017kmx, Schneck:2015eqa,
  Kang:2019fvz, Agnes:2020lzh, Adhikari:2020gxw,
  Wang:2020jxb}) then captures only a subset of all possible DM models.  Alternatively, one can limit the discussion to just a subset of possible DM interactions, for instance only to interactions that do not depend on nuclear spin, but gain in describing all possible choices for DM spin.

\subsection{Connecting to the UV}
\label{sec:connecting}

When connecting the results of direct detection experiments to the UV
theory of DM several different scales enter: the DM mass, $m_\chi$,
the scale of the DM-SM mediators, $\Lambda$, and, finally, the
standard model (SM) scales -- the masses of the SM particles and the
scale of strong interactions, $\Lambda_{\rm QCD}$.  The hierarchy
between these scales determines which EFTs constitute the tower that
connects the direct detection and UV scales, see, {\it e.g.},
Fig.~\ref{fig:eft-tower}.
 The main goal is to provide leading-order predictions for
direct detection rates for any choice of a UV theory,  which requires
constructing the EFTs for each self-consistent ordering of the EFT
scales, as well as for various DM spins and electroweak quantum
numbers.

\begin{figure}[t]\centering
	\includegraphics[width=0.49\textwidth]{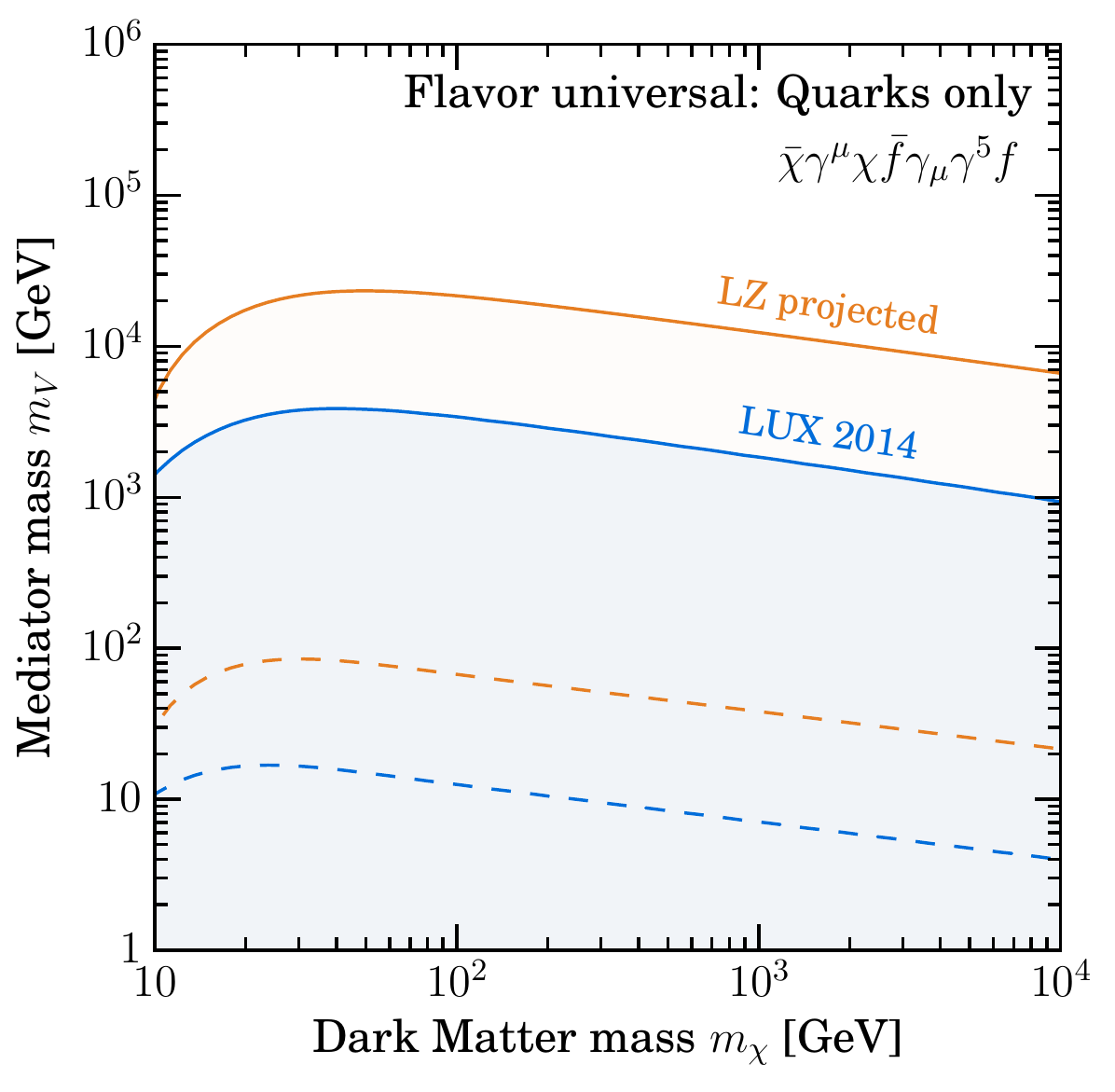}
	\includegraphics[width=0.49\textwidth]{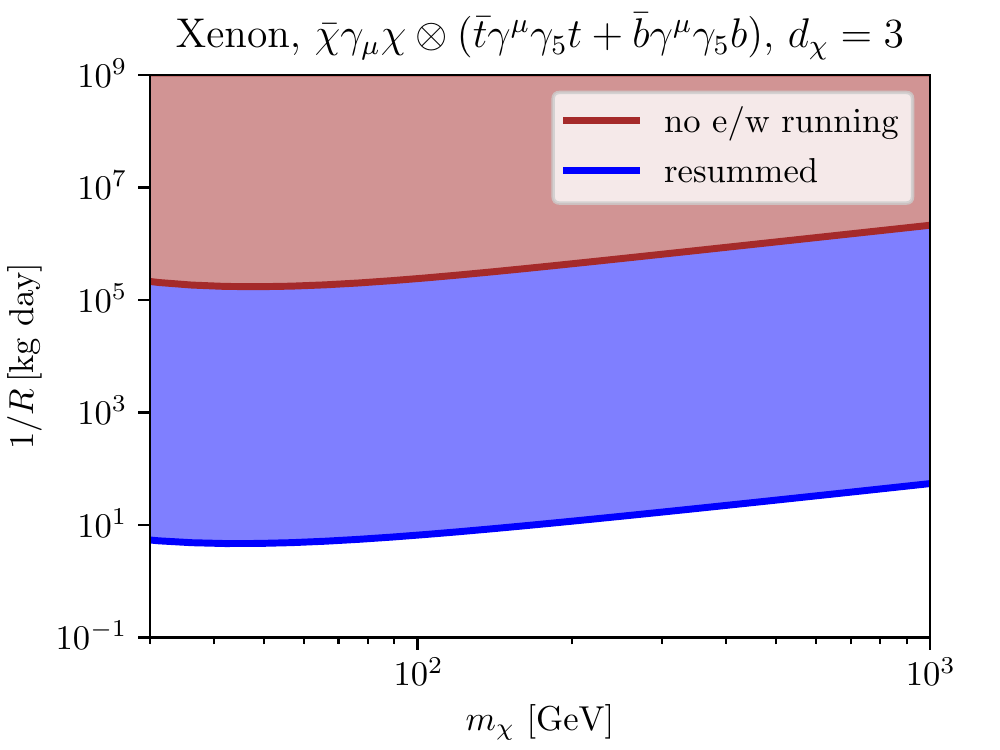}
	\caption{Left: exclusions on DM mass for $V\otimes A$ flavor universal ${\mathcal O}(1)$ couplings to quarks for  electroweak singlet Dirac fermion DM with dashed (solid) lines denoting exclusions without (with) inclusion of RG running. Right: predicted inverse rates for $V\otimes A$ operators with couplings of electroweak triplet Dirac fermion DM to only third generation quarks where resummed RGEs are included (blue) or not (red). Plots reproduced from \cite{DEramo:2016gos,Bishara:2018vix}.
		}
	\label{fig:RGE}
\end{figure}

We emphasize that in particular cases electroweak corrections need
to be included to obtain the leading predictions, since they mix
operators with very different non-relativistic limits. Fig. \ref{fig:RGE} shows the effect of electroweak corrections for two examples of  Dirac fermion DM interactions of the type $(\bar \chi \gamma_\mu \chi) (\bar u_i \gamma^\mu \gamma_5 u_i + \bar d_i \gamma^\mu \gamma_5 d_i)$, with $i=1,2,3$ the generational index, and with DM an electroweak singlet (triplet) in Fig.~\ref{fig:RGE} left (right). Fig.~\ref{fig:RGE} left assumes flavor universality. Without electroweak radiative corrections DM scattering is due to an  interaction term 
$(\bar \chi \gamma_\mu \chi) (\bar u \gamma^\mu \gamma_5 u + \bar d
\gamma^\mu \gamma_5 d)$ with very small, both spin and velocity
suppressed, nuclear matrix elements (dashed lines). At one loop there is a top yukawa induced $V\otimes V$ operator proportional to $(\alpha_t/(4\pi)) \log(M_W/\Lambda)$ that results in much larger coherently enhanced rates and more stringent exclusions (solid lines).  Fig. \ref{fig:RGE} left assumes the dimension 6 operator involves only couplings to the third generation in the UV.  For $\Lambda=1$\,TeV  the radiatively induced corrections from higher dimension $V\otimes A$ operator (blue) dominate by orders of magnitude compared to the electroweak renormalizable interactions that enter at two loops (red). 
Another illustrative example is electroweak triplet Dirac fermion DM with $V\otimes A$ couplings to only the first generation of quarks. The RG flow generates the $(\bar \chi \gamma_\mu \chi) (\bar u
\gamma^\mu \gamma_5 u + \bar d \gamma^\mu \gamma_5 d)$ interactions that is
 enhanced by a {\em quadratic} logarithm $\alpha_2^2 /(4\pi)^2 \log^2
(M_W/\Lambda)$. That is, even though the effect arises at {\em second order} in
operator mixing, the induced interaction gives the
leading contribution to the scattering rate, larger than the original
interaction by up to two orders of magnitude for scattering on heavy
nuclei. While the effect would correspond to a two-loop correction in
the ``full theory'', the DM EFT automatically captures the
leading-logarithmic part of it. A full discussion of all radiative
corrections relevant for dimension-six interactions can be found in
Ref.~\cite{Bishara:2018vix}, see also Refs.~\cite{Brod:2018ust,Crivellin:2014qxa,DEramo:2016gos,DEramo:2014nmf,DEramo:2017zqw}.

\section{EFTs for Table-top Direct Detection Experiments}
\label{sec:tabletop}

Theoretical developments over the last decade have led to the realization that compelling models of DM may have masses well below the weak scale. 
Meanwhile, new table-top experiments targeting sub-GeV DM have been proposed and are under active development. 
Sensitivity to lighter DM requires both sensors with better energy resolution and qualitatively new ideas.  
Conventional searches based on nuclear recoils have limitations: once the mass of DM drops below that of the nucleus, the detection rate suffers from a kinematic suppression. 
We know, however, that condensed matter systems host a range of small-gap excitations as well as gapless modes.
They could enable efficient extraction of a large fraction of the DM's kinetic energy even for DM much lighter than a GeV. 
See Ref.~\cite{Kahn:2021ttr} for a recent review.

To bridge the condensed matter knowledge with DM detection ideas, calculations of material responses to DM interactions are needed, with a combination of analytic and numerical tools.
In particular, EFT methods~\cite{Caputo:2019cyg,Caputo:2019xum,Catena:2019gfa,Trickle:2020oki,Catena:2021qsr,Mitridate:2021ctr} are useful both for classifying DM interactions and identifying condensed matter systems and excitations with favorable response to each type of interaction, and for generally formulating calculations of DM detection rates to facilitate automation.

For example, electronic excitations in noble gas atoms and dielectric crystals can arise from DM-electron scattering or absorption of bosonic DM coupling to electrons, and are being actively searched for by several experimental collaborations, including XENON, SuperCDMS, SENSEI, DAMIC, EDELWEISS. 
For general DM models, we can utilize the EFT framework to make signal rate predictions in terms of a set of atomic/crystal response functions, which are calculated from matrix elements of NR effective operators for DM-electron interactions in the detector medium~\cite{Catena:2019gfa,Catena:2021qsr,Mitridate:2021ctr}.

Direct detection can also proceed via production of collective excitations, such as phonons and magnons, in the primary DM scattering process. 
To calculate the rate from the set of NR effective operators (which form an expanded basis compared to the nuclear recoil case due to in-medium violation of Galilean invariance), we match them onto lattice degrees of freedom in the long wavelength limit, which include the number of $\psi=p,n,e$ particles contained in each ion $\langle N_\psi\rangle$, their total spin $\langle \vect{S}_\psi\rangle$, orbital angular momentum $\langle \vect{L}_\psi\rangle$ and tensorial spin-orbit coupling $\langle \vect{L}_\psi\otimes\vect{S}_\psi\rangle$. 
These four types of couplings are referred to as crystal responses, as they play an analogous role as nuclear response functions in the nuclear recoil EFT calculation.
In the present case, these crystal responses enter the effective DM-lattice scattering potential, which is then quantized in terms of phonon or magnon modes. 
In the simplest cases, phonon excitations in a crystal proceed through $\langle N_\psi\rangle$~\cite{Knapen:2017ekk,Griffin:2018bjn,Trickle:2019nya,Griffin:2019mvc} and magnon excitations proceed through $\langle \vect{S}_e\rangle$~\cite{Trickle:2019ovy}. 
More generally, all four types of crystal responses can lead to phonon excitations in appropriately chosen targets, while  both $\langle \vect{S}_e\rangle$ and $\langle \vect{L}_e\rangle$ can lead to magnon excitations. 

This EFT framework for direct detection with collective excitations~\cite{Trickle:2020oki} has been implemented in a publicly available code \textsf{PhonoDark}~\cite{phonodark}, and provides theory support for the ongoing experimental effort on phonon readout via {\it e.g.}\ transition edge sensors in SPICE and HeRALD experiments~\cite{tesseract}. 
It also sets up the stage for investigating other condensed matter systems to identify new detector targets, a direction we look forward to further pursuing in the future.

\section{Simplified mediator models for colliders}
\label{sec:colliders}

We have not yet observed any signals of DM being produced at colliders. This could be due to DM being too weakly coupled to the SM or simply being too heavy. An interesting intermediate possibility is that DM is light enough to be produced at colliders, and has appreciable couplings to other states, but interacts with the visible sector only through mediators that are too heavy to be efficiently produced at the LHC. At low energies  the DM interactions with the SM are then described by an appropriate EFT, where the heavy mediators between DM and the SM particles were integrated out.
Such an EFT based approach is ideal for interpreting the results of DM scattering in direct detection experiments, in which the typical momentum exchange, $q\lesssim 200$ MeV, is much less than the mediator mass in most DM models, see Sec.~\ref{sec:direct:detection}. 

The use of such EFTs to describe the results of LHC searches for DM, on the other hand, is more suspect. 
The cross sections for DM production, induced by non-renormalizable operators, grow with energy. 
For example, a dimension-6 operator operator of the form $(\bar \chi \gamma_\mu \chi)(\bar q \gamma^\mu q)$ is suppressed by $1/\Lambda^2$, where $\Lambda$ is the NP scale. This then results in a DM production cross section that scales as $\sigma \sim E^2/\Lambda^4$,  and thus a large part of experimental sensitivity at the LHC may well come from collision energies at which the EFT description breaks down. There is an interplay between how precisely the SM backgrounds are predicted and how much statistics the LHC experiments are able to gather, which then translates to bounds on $\Lambda$. Typically, these are not stringent enough to be in the EFT regime, $\Lambda \gg E$, except for a small region of parameter space where the mediator couplings are close to the nonperturbative limit. 
In a very large part of the relevant parameter space the mediators are instead light enough to be produced directly in $pp$ collisions. It is thus much better to interpret the bounds on DM production at the LHC  by constructing the EFTs where the SM is supplemented by both DM and the mediator -- the so called ``simplified models'' \cite{Frandsen:2011cg,Agrawal:2011ze,Agrawal:2010fh,Goodman:2011jq,An:2012va,Frandsen:2012rk,Garny:2012eb,Bell:2012rg,An:2012ue,Chang:2013oia,An:2013xka,Bai:2013iqa,DiFranzo:2013vra,Alves:2013tqa,Arcadi:2013qia,Bai:2014osa,Lebedev:2014bba,Abdallah:2014hon,Buckley:2014fba,Alves:2015pea,Berlin:2015wwa,Ismail:2016tod,Alves:2016cqf,Arcadi:2017jqd,Evans:2017kti,Alanne:2017oqj,Alanne:2020xcb,Argyropoulos:2022ezr}. The complete UV models of DM would contain further degrees of freedom, which are assumed to be heavy and integrated out. 

The simplified models differ in the assumed spin and the SM gauge quantum numbers of the mediators. The most commonly considered simplified models are: the $s$-channel color-neutral vector or axial-vector mediator, the  $s$-channel color-neutral scalar or pseudo-scalar mediator, and the $t$-channel color-triplet scalar or pseudo-scalar mediator \cite{Abdallah:2014hon,Abdallah:2015ter,Boveia:2016mrp,Abercrombie:2015wmb,Albert:2016osu,Albert:2017onk,LHCDarkMatterWorkingGroup:2018ufk}. It is important that the simplified models satisfy the SM gauge invariance in order not to arrive at spurious violations of unitarity in predictions for the cross sections. In some cases this requires introducing more states beyond just the mediator.  For instance, for a pseudo-scalar mediator the community has endorsed the use of the simplest such choice: a two Higgs doublet with a singlet pseudo-scalar that couples to DM (the 2HDM+$a$ model). In some instances it is also important to include electroweak radiative corrections when comparing the results of LHC searches with direct detection results, see Section \ref{sec:connecting}.

Phenomenologically, there are several qualitative differences between the use of EFTs with and without mediators. 
Importantly, in the EFTs that include the mediators one is able to capture the constraints on the UV models of DM which are placed by searches for mediators when these decay not to DM pairs but to other visible final states such as muons, electrons, jets, etc. In many cases these are the most stringent constraints on the simplified models that are otherwise missed in EFTs constructed by integrating out the mediators. In simplified models one still needs to make a number of choices for mediator couplings when interpreting the data (for instance, the flavor composition of couplings to quarks, whether or not the mediators couple to leptons, etc). 
While simplified models assume minimal field content, the freedom of choice for various couplings results in simplified models that are not so very simple and thus capture many of the most important features of the full UV models.

\section{EFTs for DM self interactions}
\label{sec:sidm}

The Sommerfeld effect discussed in Sec.~\ref{sec:somm} can be relevant not only for dark matter annihilation, but also for dark matter scattering. Self-interacting  dark matter has received a great deal of attention, in part due to its possible role in explaining the  structure of galaxies on smaller scales; see~\cite{Tulin:2017ara} for an extensive review. It is well-known that relatively long-range forces, like Yukawa interactions mediated by a scalar or vector much lighter than the dark matter particle itself, can lead to enhanced scattering at low velocity, which may be understood by  resumming ladder diagrams in QFT or  by solving a non-relativistic Schr\"odinger problem. For various studies of the Sommerfeld effect for Yukawa interactions, see~\cite{Arkani-Hamed:2008hhe,Pospelov:2008jd,Buckley:2009in,Feng:2009hw,Loeb:2010gj,Tulin:2013teo}.

Because self-interactions can involve important non-relativistic effects, it is appropriate to match the underlying relativistic QFT of self-interacting dark matter onto a non-relativistic effective theory~\cite{Caswell:1985ui,Lepage:1997cs}. This was first done for dark matter self-interactions in~\cite{Bellazzini:2013foa}, using the same classification of operators~\cite{Dobrescu:2006au} that underlies the EFT for DM--nucleus scattering discussed above. In the initial work along these lines, a detailed matching to the underlying QFT was not carried out, which effectively meant that computations were being done in potentials with additional (possibly strong) short-range scattering---an artifact  of improper matching, rather than actual physics. In particular, different early studies reached different conclusions about whether dark matter scattering via the exchange of a pseudoscalar mediator exhibits Sommerfeld enhancement. Because  such  models can  arise in natural  ways ({\it e.g.}, with the mediator being a pseudo-Nambu-Goldstone boson), it is important to assess whether the  Sommerfeld enhancement  exists  in this case and what the correct velocity-dependence of pseudoscalar-mediated  DM scattering is.

\begin{figure}[t]\centering
\includegraphics[scale=0.5]{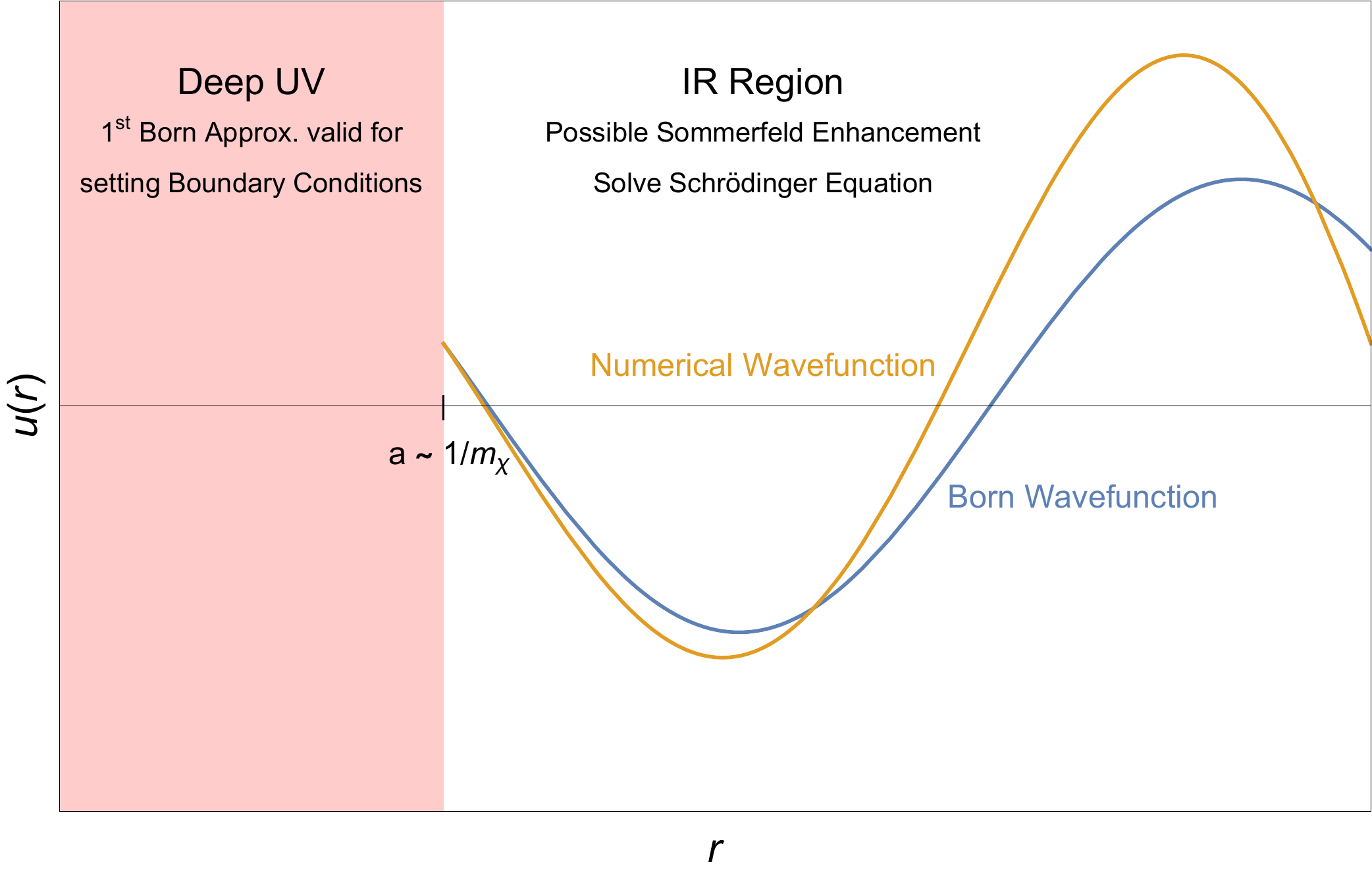}
\caption{Illustration of the matching  procedure from QFT to a non-relativistic Schr\"odinger problem with boundary at small radius, reproduced from~Ref.~\cite{Agrawal:2020lea}. }
\label{fig:SIDMmatching}
\end{figure}

Recently, these calculations have been clarified via a simple new procedure for correctly matching the predictions of relativistic QFT onto a non-relativistic Schr\"odinger problem. The key idea is illustrated in Fig.~\ref{fig:SIDMmatching} (reproduced from Ref.~\cite{Agrawal:2020lea}). The nonrelativistic effective theory is expected to break down at high momentum, and hence at short distances. This makes possible a matching procedure where the non-relativistic EFT is defined  only outside a finite radius, {\it e.g.}, $a \sim 1/m_\chi$. The tree-level QFT cross section at low velocities is matched to the leading order Born approximation in non-relativistic QM. This matching can be achieved by imposing a boundary condition on the wavefunction at the radius $a$. Subsequently, the full nonperturbative solution in the non-relativistic EFT can be achieved simply by solving the Schr\"odinger equation at $r > a$ with the correct boundary condition. The boundary condition essentially arises from evaluating the Born approximation to the wave function in the region $r < a$, which turns out to be finite for the effective potentials arising from tree-level QFT (see also~\cite{Parikh:2020ggm}). With this matching procedure, one can explicitly check that dark matter scattering via a pseudoscalar mediator does not exhibit any  Sommerfeld enhancement: the tree-level QFT calculation is a good approximation to the solution to the Schr\"odinger equation. This can also be understood by a study of the behavior of ladder diagrams, and reinforces earlier claims of no Sommerfeld enhancement for pseudoscalar mediators~\cite{Kahlhoefer:2017umn}. The new matching procedure can be applied to a variety of other problems, including dark matter annihilation. Its extension beyond tree level would be an interesting target of further study.

\section{Conclusions}

The landscape of phenomenologically viable dark matter models is vast.  It is often the case that there can be a large separation of scales involved in relating such models to observables, implying that effective field theory techniques could be relevant.  This could be due to the parameters of the theory, \emph{e.g.}~if the dark matter mass is much larger than the electroweak scale.  It could also be the result of the kinematics relevant to the observable of interest, \emph{e.g.}~direct detection experiments are looking for non-relativistic collisions between dark matter and nucleons.

The EFT paradigm provides a systematic approach for computing observables as a low energy expansion.  When viewed from the bottom up, it is a tool for characterizing the allowed interactions among the propagating degrees of freedom.  EFT is also useful from the top down: in the presence of a large separation of scales, one can integrate out the heavy physics to derive an EFT Lagrangian which often simplifies calculations and can often facilitate more sophisticated calculations that could not have been done using the UV theory along.  In particular, one can utilize renormalization group evolution to resum large logarithms that appear when one calculates using the UV theory directly.  Both of these themes play a prominent roles in the work described above.

Clearly, the importance of applying EFT techniques to the DM question has led to significant improvements in our ability to calculate processes that are relevant across a huge range of experiments and observations.  It additionally allows us to systematically classify the possible types of interactions the DM can have with the Standard Model, so that we can be systematic in our experimental approach.  Further refinements and new ideas for EFT applications in DM are an active area of research.  EFT will continue to play a pivotal role in the search for an experimental signature of DM.

\vspace{0.2in}

{\bf Acknowledgements.} M.B.~is supported by the DOE (HEP) Award DE-SC0019470.
A.L.F. is supported in part by the US Department of Energy Office of Science under Award Number DE-SC0015845.
J.Z. acknowledges support in part by the DOE grant de-sc0011784 and NSF OAC-2103889. 
M.R.~is supported by the DOE Grant DE-SC0013607, the NASA Grant 80NSSC20K0506, and the Alfred P.~Sloan Foundation Grant No.~G-2019-12504.
R.S. acknowledge support by the United States Department of Energy under Grant Contract No.~DE-SC0012704.
T.C.~is supported by the U.S. Department of Energy, under grant number DE-SC0011640. 
Z.Z.\ is supported by the U.S.\ Department of Energy under the grant DE-SC0011702. M.P.S.~is grateful to the Mani L. Bhaumik Institute for Theoretical Physics for support.
  
\bibliographystyle{JHEP}

\bibliography{DMEFT}

\end{document}